# Navigating Knowledge: Patterns and Insights from Wikipedia Consumption

Tiziano Piccardi[1] (0000-0001-9314-1440),
Robert West[2] (0000-0002-3984-1232)

[1] Computer Science Department, Stanford University, Stanford, CA, USA
[2] School of Computer and Communication Sciences, EPFL, Lausanne, Switzerland

**Abstract**
The Web has drastically simplified our access to knowledge and learning, and fact-checking online resources has become a part of our daily routine. Studying online knowledge consumption is thus critical for understanding human behavior and informing the design of future platforms. In this Chapter, we approach this subject by describing the navigation patterns of the readers of Wikipedia, the world's largest platform for open knowledge. We provide a comprehensive overview of what is known about the three steps that characterize navigation on Wikipedia: (1) how readers reach the platform, (2) how readers navigate the platform, and (3) how readers leave the platform. Finally, we discuss open problems and opportunities for future research in this field.

## 1. Introduction

Evolution has equipped humans with a remarkable capacity for knowledge-seeking, and humans have continually shaped the world around them to improve access to knowledge. Many of the most consequential evolutionary, cultural, and technological advances have enhanced our ability to find, ingest, process, and transfer knowledge. From the development of language and writing systems to modern telecommunication channels, we have persistently sought to push the boundaries of knowledge sharing.

In our history, as part of this constant effort of sharing knowledge, encyclopedias have played a crucial role. Since antiquity, humans have developed ways to keep track of and share, what we know about the world. From the ancient Pliny's Naturalis Historia, which served as an editorial model, to the development of the modern concept of an encyclopedia in Enlightenment France, this effort in collecting knowledge served the same ideal. In the 18th century, philosopher Denis Diderot (Diderot, 1876) defined encyclopedias as a way to disseminate knowledge to people who live with us or will come after us, in a virtuous cycle "so that the work of preceding centuries will not become useless to the centuries to come". Fast-forward to the 20th century, with tremendous technological progress, the way we think about accessing knowledge changed drastically. In 1945, Vannevar Bush (Bush & others, 1945) sketched his vision of an information management device—the Memex—that would allow users to retrieve information quickly and enhance their memory by interlinking documents following the associations in the human mind. In the last decades, digitalization has brought us closer to his visionary idea. From the availability of expert-curated encyclopedias like Microsoft Encarta to the development of online crowdsourced Web encyclopedias like Wikipedia, our access to knowledge has become ubiquitous and effortless.

Given the central importance of knowledge seeking to human nature—epitomized by the view of humans as informavores (Machlup, 1983)—, understanding how humans seek information is of key significance across disciplines, both in the basic and applied sciences. In the basic sciences, biologists, psychologists, and anthropologists, among others, stand to gain fundamental insights into how humans function, whereas, in the applied sciences, such insights can enable the design of more effective tools and information environments, such that humans can more readily find relevant knowledge in an ever-surging flood of information.

To address the fundamental question of how humans interact with knowledge, Wikipedia represents an invaluable resource. Besides simplifying and democratizing access to knowledge, Wikipedia represents the ideal candidate to investigate human behavior around knowledge. Thanks to its wide adoption and a rich network of concepts that people can navigate and interact with, researchers can finally collect data that gives us an unprecedented view into the nature of human knowledge-seeking processes. These insights not only deepen our understanding of how people navigate and engage with information but also have significant implications for designing web platforms that are more effective and aligned with diverse user needs.

This Chapter summarizes research with the goal of understanding how we interact with knowledge. We focus on how readers navigate Wikipedia in English, Wikipedia's largest language edition. We organize our characterization around three moments of the dynamic of knowledge consumption on Wikipedia: how readers reach the platform, how they navigate the content, and how they leave the websites.

Data sources

Collecting the data to investigate how readers move across online content poses significant challenges. Surveys (Younger, 2010) and thinking-out-loud studies (Muramatsu & Pratt, 2001) are prone to cognitive biases, as humans generally perform poorly at introspection (Nisbett

& Wilson, 1977). At the same time, lab-based experiments typically involve small samples consisting of biased populations (e.g., university students). They are thus frequently not representative and might lack statistical power. The ideal approach to observing the natural navigation patterns of the readers, without altering their behavior, is to collect their activities passively. Like many online services, Wikipedia stores these activities for analytics purposes (for a limited time) in its server logs and can offer researchers valuable insights into the readers' behavior. These activities can be aggregated by combining the requests of the same client to investigate the natural navigation of the readers on Wikipedia.

The downside of working with raw server logs is the need for privileged access to sensitive information such as IPs and geo-locations that, for privacy reasons, should be granted with care. To promote research in this direction and overcome this limitation, the Wikimedia Foundation releases with monthly frequency the public clickstream (Wulczyn & Taraborelli, 2015) for multiple language editions. This dataset contains transition counts for pairs of articles, giving researchers a practical way to investigate how readers move from document to document. The clickstream has also been used to generate synthetic navigation paths through biased random walks (Rodi, Loreto, & Tria, 2017) that select the next article according to the clickstream probability. Although the clickstream is an aggregated and filtered version of the server logs that preserves the readers' privacy, it has been shown to approximate real navigation with a good level of accuracy (Arora, Gerlach, Piccardi, García-Durán, & West, 2022; Piccardi, Gerlach, Arora, & West, A Large-Scale Characterization of How Readers Browse Wikipedia, 2023). On common tasks such as link prediction and topic similarity computation, the conclusions obtained using the clickstream are qualitatively the same as for the private data (despite small, statistically significant differences).

An alternative approach to characterizing human navigation relies on digital traces obtained via navigation games such as Wikispeedia (West, Pineau, & Precup, Wikispeedia: An Online Game for Inferring Semantic Distances between Concepts, 2009) and TheWikiGame. They offer a convenient way to collect data, providing users entertainment by asking them to start from a random article and reach a target page with as few clicks as possible through internal links. The trajectories are then collected as sequences and reveal how people move across Wikipedia when they look for specific content. In contrast to natural navigation, these trajectories, denoted as targeted navigation, come with an unambiguous definition of success (i.e., reaching the target article). They let researchers determine if the navigation is terminated and study the strategies used by players to traverse the information network.

However, although the clickstream and navigation games represent excellent resources for approximating the behavior of Wikipedia readers, they cannot replace the server logs entirely. The fine-grained private activities stored on the servers remain an irreplaceable resource for studies that focus on the situated, sequential behavior of readers in a natural setup. The logs include timestamps, geolocations useful to convert the timestamp in local time, and user identifiers that can help us answer questions that rely on modeling navigation trails of the same user across different articles. Sections 3 and 4 of this Chapter are largely based on observations obtained thanks to logs collected on the server (Piccardi, Gerlach, Arora, & West, A Large-Scale Characterization of How Readers Browse Wikipedia, 2023; Piccardi, Gerlach, Arora, & West, Going down the Wikipedia Rabbit Hole: Characterizing the Long Tail of Reading Sessions, 2022), whereas Section 5, which describes the outgoing traffic, relies on logs collected on the client side (Piccardi, Redi, Colavizza, & West, Quantifying engagement with citations on Wikipedia, 2020; Piccardi, Redi, Colavizza, & West, On the Value of Wikipedia as a Gateway to the Web, 2021).

Readers' motivations and content popularity

There are a variety of different factors that motivate readers to consume Wikipedia content. Surveys show that some of these reasons include learning more about current events, media coverage of a topic, personal curiosity, work or school assignments, or boredom (Singer, et al., 2017). The motivation that brought the reader to Wikipedia is also associated with differences in reading behavior. For instance, people exploring Wikipedia out of boredom tend to have long sessions with fast transitions between articles, whereas readers interested in learning about a subject spend more time on a few relevant pages. Here, the behaviors observed in the different language editions highlight some sociocultural differences. For instance, Wikipedia plays a particularly important role as a source of knowledge for countries with a low Human Development Index, with readers from these countries exhibiting more in-depth reading behavior (Lemmerich, Sáez-Trumper, West, & Zia, 2019) and spending more time reading the content of the visited articles (TeBlunthuis, Bayer, & Vasileva, 2019). Reading behavior also varies across genders (Johnson, et al., 2020).

Another important driver of navigation is curiosity. Studies—conducted both in-lab and through server logs of the mobile app—investigating the link network explored by readers reveal that their behavior aligns with curiosity-driven information-seeking patterns (Lydon-Staley, Zhou, Blevins, Zurn, & Bassett, 2021; Zhou, et al., 2024). These patterns, borrowing terminology from philosophy, are referred to as "hunters" and "busybodies."

Content consumption also is affected by readers' preferences. Different reasons to consume the content combined with article properties show patterns (Lemmerich, Sáez-Trumper, West, & Zia, 2019) associated with four categories of behaviors described as exploration, focus, trending, and passing. At the same time, not all articles receive the same attention. Readers show specific preferences with respect to types of content. Documents about entertainment (Spoerri, 2007), particularly music and TV shows (Waller, 2011), receive more attention, and engagement can even predict box office success (Mestyán, Yasseri, & Kertész, 2013). At the same time, readers also exhibit preferences in the type of images on the page. Readers click more on images associated with articles about visual arts, transports, and biographies of less well-known people (Rama, Piccardi, Redi, & Schifanella, 2021).

However, these preferences are not shared by everyone. Different population segments have diverse information needs (Waller, 2011), and the ratio of the topics consumed varies based on their lifestyle. The skewness in attention among different people and for different content creates a gap between production and consumption: the most popular articles are not always the most edited (Lehmann, Müller-Birn, Laniado, Lalmas, & Kaltenbrunner, 2014).

Additionally, the preferences of the readers are not always stable in time. Wikipedia articles can experience sequences of bursts of attention caused by external factors (Ratkiewicz, Fortunato, Flammini, Menczer, & Vespignani, 2010), such as an Academy Award nomination, or change their popularity after updates to the page design, such as the introduction of new features, including link preview (Chelsy Xie, Johnson, & Gomez, 2019). Similarly, during the COVID-19 pandemic, exogenous factors such as mobility restrictions impacted the type of content people sought on Wikipedia. Topics associated with self-actualization and possibly associated with more time spent at home, such as entertainment and video games, experienced increased attention (Ribeiro, et al., 2020). Wikipedia is a common destination for learning about current events, with readers shifting their attention across clusters of interconnected articles as they follow evolving topics (Gildersleve, Lambiotte, & Yasseri, Between news and history: identifying networked topics of collective attention on Wikipedia, 2023). However, not all external factors impact content popularity. For instance, the awareness campaign to promote Wikipedia in Hindi (Chelsy Xie, Johnson, & Gomez, 2019) showed no significant change in traffic.

Besides individual bursts of popularity for specific articles, Wikipedia has regular access patterns that follow the natural human circadian rhythm (Piccardi, Gerlach, & West, Curious Rhythms: Temporal Regularities of Wikipedia Consumption , 2024). Server logs (Reinoso, Gonzalez-Barahona, Robles, & Ortega, 2009) highlight the daily and weekly patterns and how this data could be exploited to provide valuable insights into the readers' behavior and design an efficient and scalable infrastructure adapted to reading patterns (Reinoso, Munoz-Mansilla, Herraiz, & Ortega, 2012; Piccardi, Gerlach, Arora, & West, A Large-Scale Characterization of How Readers Browse Wikipedia, 2023; Piccardi, Gerlach, Arora, & West, Going down the Wikipedia Rabbit Hole: Characterizing the Long Tail of Reading Sessions, 2022).

## 2. How readers reach Wikipedia

Wikipedia's traffic is influenced by its connections to the larger Web ecosystem and its interdependence with external platforms. Each request to Wikipedia servers leaves a digital trace in the server logs, identifying the traffic's origin. This data enables the distinction between external and internal traffic, offering valuable insights into how and when people access Wikipedia (Piccardi, Gerlach, Arora, & West, A Large-Scale Characterization of How Readers Browse Wikipedia, 2023).

Traffic origin

Besides explicitly typing the URL into the browser, people can reach Wikipedia's articles in various ways, such as clicking links on external websites, social media, or search engine result pages.

**Search engines** are responsible for most of the incoming traffic received by Wikipedia, representing the preferred way to access its content. Almost 78% of the incoming traffic to Wikipedia originated from search engines, but Wikipedia's relationship with search engines is deeper than just traffic acquisition. Google, the most common external origin, shows a mutual dependency with Wikipedia (McMahon, Johnson, & Hecht , 2017)). Wikipedia depends largely on Google for its readership, but at the same time, Wikipedia contributes to Google's success by answering a large portion of the queries posed by its users. Indeed, Google returns Wikipedia articles on the first page of results for 67%-85% of the queries, depending on the type of searched content (Vincent & Hecht, A Deeper Investigation of the Importance of

Wikipedia Links to Search Engine Results, 2021; Laurent & Vickers, 2009). The interdependency between search engines and Wikipedia also results in a high correlation between the volume of searches for one term and page-loads of the relative article. This association means that it is possible to partially reconstruct search engines' activities by investigating the incoming traffic on Wikipedia (Yoshida, Arase, Tsunoda, & Yamamoto, 2015). There is, however, a critical tradeoff in the relationship between Wikipedia and search engines. On the one hand, Wikipedia's content improves search results, for example, via knowledge panels and AI-based summaries; on the other hand, this might keep users whose information needs are already satisfied by the knowledge panel from visiting Wikipedia itself.

**External websites**, such as social media and Q&A platforms, are another source of incoming traffic for Wikipedia, accounting for 1.5% of the external requests. Among those, the most common sources are Facebook (15.6%), Reddit (9.6%), YouTube (8.0%), and Twitter (4.3%). Like search engines, these external websites have a strong dependency on Wikipedia. Wikipedia is a common domain in the links posted on platforms such as StackOverflow and Reddit (Gómez, Cleary, & Singer, 2013; Vincent, Johnson, & Hecht, Examining Wikipedia with a broader lens: Quantifying the value of Wikipedia's relationships with other large-scale online communities, 2018), and their community engages more with posts containing links to Wikipedia articles. The benefit offered in the other direction, from these platforms to Wikipedia, is less obvious. The presence of links on external platforms influences the attention received by Wikipedia in terms of page-load counts (Moyer, Carson, Dye, Carson, & Goldbaum, 2015). However, this interest does not last, and is not translated into measurable actions, with Wikipedia not recording an increase of edits (Vincent, Johnson, & Hecht, Examining Wikipedia with a broader lens: Quantifying the value of Wikipedia's relationships with other large-scale online communities, 2018; Morgan & Johnson, 2020) for the linked articles.

**Unspecified origin** is another frequent type of incoming traffic, covering around 20% of the requests. Multiple reasons can produce a request without an explicit origin, including direct access via the browser history, redirects from apps, bookmarks, search toolbars, or when the link source has explicitly turned on the "noreferrer" property. In particular, direct access by using the browser history could be associated with a very common revisiting pattern described in many studies about the general behavior of users on the Web (Tauscher & Greenberg, 1997; Anderson, Kumar, Tomkins, & Vassilvitskii, 2014). In around 18% of the pageloads, a reader loads the same document at least twice in the following month, and 11% of two subsequent pageloads happening within 1 hour are composed of the same article without other Wikipedia content consumed in between.

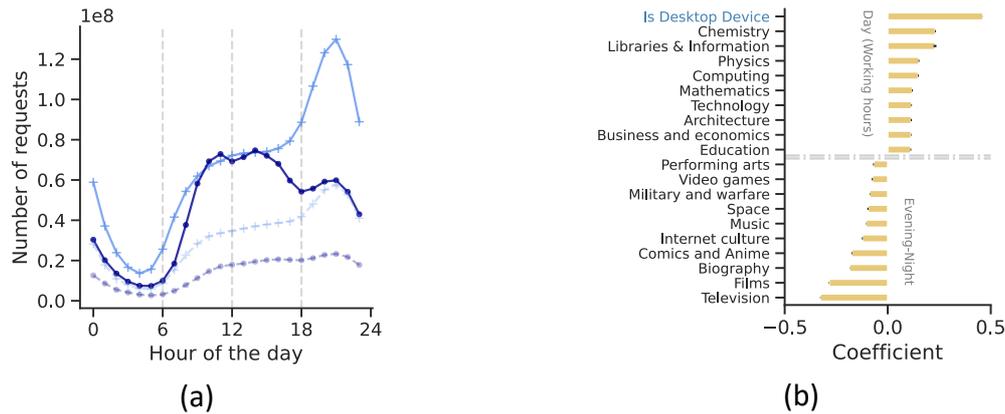

*Figure 1. (a) Average incoming traffic to Wikipedia by time of the day and device. (b) Feature contributions to the logistic model predicting if the requests arrived during working hours. (Figures borrowed from* (Piccardi, Gerlach, Arora, & West, A Large-Scale Characterization of How Readers Browse Wikipedia, 2023)*)*

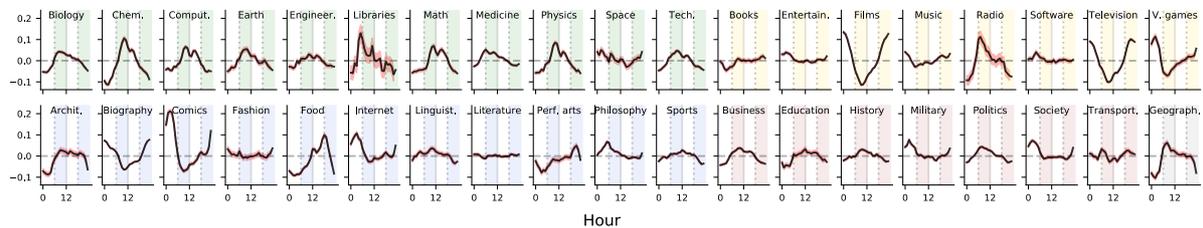

*Figure 2. Linear regression coefficients of the interaction between topics and hour, representing the variation in pageloads relative to the general average* (Piccardi, Gerlach, & West, Curious Rhythms: Temporal Regularities of Wikipedia Consumption , 2024). *These coefficients isolate the contribution of the topic, disentangling the effects of country and device type (mobile vs. desktop). Background colors indicate top-level topics: STEM (green), MEDIA (yellow), CULTURE (blue), HISTORY & SOCIETY (red), and GEOGRAPHICAL (gray). Specific topics within each top-level category are arranged alphabetically. Red bands denote 95% confidence intervals.*

Temporal access patterns

The temporal distribution of the incoming traffic shows a regular circadian rhythm. Figure 1a shows the hourly average incoming traffic split by access method (desktop vs. mobile) and weekday vs. weekend. Both desktop and mobile traffic show a similar pattern during the day, with a substantial increase in mobile requests in the evening. Wikipedia has lower incoming traffic during weekends, but the temporal distribution on weekends resembles that on working days. The desktop distribution shows dips at 12:00 and 18:00, mirroring work rhythms with a lunch break around noon and the end of work (and possibly commuting) in the evening.

The daily variation also affects the topics consumed by readers. Figure 1b shows the topics (obtained via the ORES library (Halfaker & Geiger, ORES: Lowering Barriers with Participatory Machine Learning in Wikipedia, 2019)) and the device type that are the most and least associated with working hours (9:00-18:00) (quantified via the coefficients of a logistic regression fit to predict in what time interval a modeled pageload occurred). During working hours, readers tend to use more desktop devices and consume more content associated with Education and STEM. In contrast, articles on entertainment—such as Television, Films, and Biographies—are strongly associated with evening and nighttime browsing. Figure 2 reinforces this observation, showing that each topic has a unique daily attention pattern, reflecting when readers typically seek the information provided on those pages

These access patterns are shaped by socio-cultural contexts and vary significantly across countries (Piccardi, Gerlach, & West, Curious Rhythms: Temporal Regularities of Wikipedia Consumption , 2024). They provide insights into global reading behaviors, illustrating that Wikipedia is not a single-themed resource but fulfills different information needs depending on the context. The same readers may turn to Wikipedia for different purposes at various times of the day.

## 3. How readers navigate Wikipedia

Wikipedia server logs allow the grouping of requests originating from the same client, offering a convenient way to study the behavior of the same reader in time.

Natural navigation

**Transition dynamics.** Readers tend to move across pages relatively quickly (Piccardi, Gerlach, Arora, & West, A Large-Scale Characterization of How Readers Browse Wikipedia, 2023). The inter-event time distribution between two consecutive page-loads of the same reader has a median of 74 seconds (63 and 93 seconds for mobile and desktop devices, respectively). However, the distribution is long-tailed, with 22% of the pairs separated by more than one hour. Additionally, readers frequently do not rely on internal links to transition between two articles, but on external pages (typically search engines) by leaving and re-entering Wikipedia. These external transitions are not rare: in around 35% of the pairs of the same reader with less than one hour between the two events, the second page was reached through external navigation. This observation is corroborated by Figure 3a, which shows that transitions via internal links are even less common than transitions via external navigation for pairs with an inter-event time greater than 3 minutes and 48 seconds. This behavior suggests that readers often leave Wikipedia and return shortly after, engaging in a surface-level exploration of content consistent with a random surfing model (Geigl, et al., 2015).

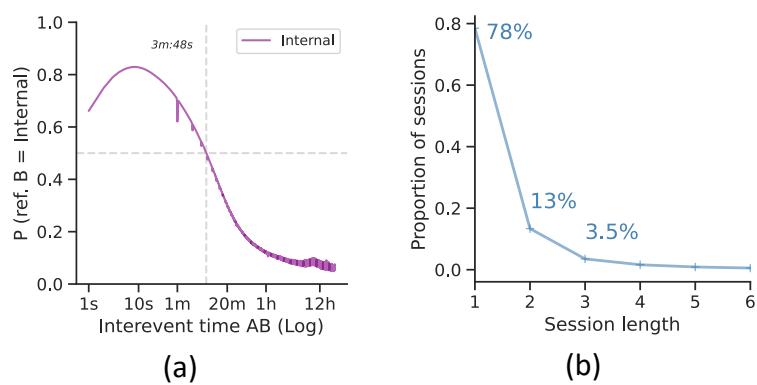

*Figure 3. (a) Probability of transition to the next page through internal navigation in function of the interevent time (b) Distribution of session length. (Figures borrowed from* (Piccardi, Gerlach, Arora, & West, A Large-Scale Characterization of How Readers Browse Wikipedia, 2023)*)*

**Traffic relayers.** Properties of the articles, such as topic and document layout, are associated with various consumption patterns. For example, the clickstream shows that articles about historical military events relay more traffic than content related to architecture (Dimitrov, Lemmerich, Flöck, & Strohmaier, 2018), and different types of articles, such as lists and disambiguation pages, relay traffic in different ways by acting as distributors of traffic (Gildersleve & Yasseri, Inspiration, Captivation, and Misdirection: Emergent Properties in Networks of Online Navigation, 2018). The different navigation strategies are also associated with the document structure. Readers tend to have a positional bias in the links clicked, preferring links positioned at the top of the page (Lamprecht, Lerman, Helic, & Strohmaier, 2017; Dimitrov, Singer, Lemmerich, & Strohmaier, 2017; Paranjape, West, Zia, & Leskovec, 2016) and navigational links in the infobox that bring the reader to the next or previous event of a series (Piccardi, Gerlach, Arora, & West, Going down the Wikipedia Rabbit Hole: Characterizing the Long Tail of Reading Sessions, 2022).

**Definition of reader sessions.** In contrast to existing pre-processing pipelines in data analysis and machine learning (e.g., tokenization and normalization steps in NLP), we still lack best practices for modeling navigation paths. Current studies rely on domain-specific heuristics, each capturing different aspects of navigation behavior. One approach that generates pathways defined as reading sequences (Piccardi, Gerlach, Arora, & West, A Large-Scale Characterization of How Readers Browse Wikipedia, 2023) represents a list of subsequent pageloads, sorted by time, with all the interevent times between the events below a threshold. The common threshold used to segment sessions is 1 hour (Halfaker, et al., 2015). This approach models how readers consume content in temporal order, but does not describe more complex dynamics, such as clicking multiple links in the same article. To capture such more complex dynamics, another approach, defined as navigation trees (Piccardi, Gerlach, Arora, & West, A Large-Scale Characterization of How Readers Browse Wikipedia, 2023; Paranjape, West, Zia, & Leskovec, 2016), connects pageloads sequentially using the HTTP referrer information. This strategy generates trees, as users can open multiple tabs from the same page and back-track in their navigation with the browser's back button and choose a different path. The resulting data structure describes how readers traverse Wikipedia by following internal links. The downside is the difficulty of incorporating time dynamics and capturing content consumption for pages not reached through internal clicks, even if close in time. This approach, limited to trees without branching, can be approximated with the public clickstream by running a biased random walker (Arora, Gerlach, Piccardi, García-Durán, & West, 2022; Rodi, Loreto, & Tria, 2017). The remainder of this section focuses on navigation trees obtained from the private server logs (not the public clickstream) and their properties.

**Average length of the reader sessions.** Most sessions consist of a single page load, but the length distribution exposes a long tail. Figure 3b summarizes the session length distribution: 78% of the navigation trees are composed of a single pageload, and the average length is 1.2 pageloads per session. The average session length varies during the day, with readers engaging in longer sessions during the evening and night. Similarly, very long sessions, often described as wiki rabbit holes, are more frequent at night (Piccardi, Gerlach, Arora, & West, Going down the Wikipedia Rabbit Hole: Characterizing the Long Tail of Reading Sessions, 2022). The length of the session is also associated with the device and the topics consumed by the reader. Longer sessions, including wiki rabbit holes, tend to start from desktop devices and entertainment articles. On the other hand, readers tend to stop at the first page in sessions starting with an article associated with STEM or medicine. Figure 4a shows the properties of the pageload that are most and least predictive of continuing the navigation through internal clicks after the first article.

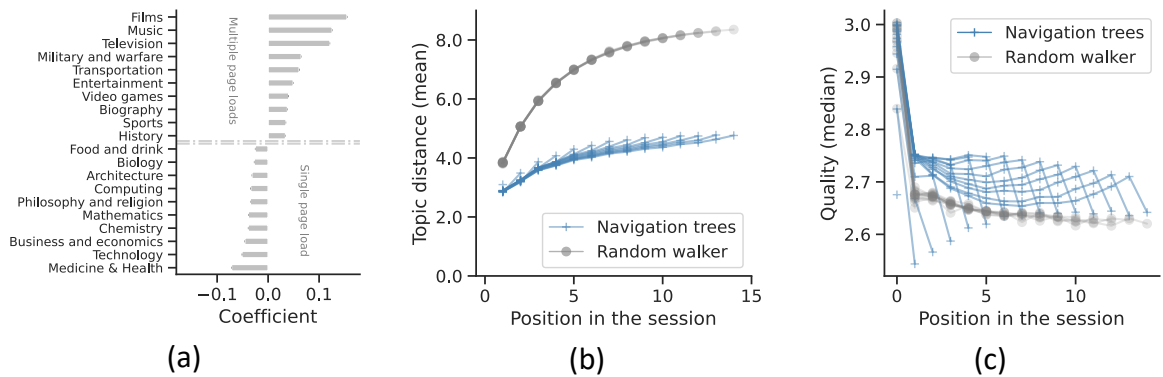

*Figure 4. (a) Feature contributions to the logistic model predicting if the reader will continue the navigation beyond the first page. (b) Within-session evolution of topic distance from first article. (c) Within-session evolution of articles' quality stratified by session length. (Figures borrowed from (Piccardi, Gerlach, Arora, & West, A Large-Scale Characterization of How Readers Browse Wikipedia, 2023))*

**Evolution of readers' sessions.** Even if the internal navigation tends to be short, given the popularity of Wikipedia, readers leave in the server logs traces of millions of long sessions every month. Understanding the evolution of the session beyond the first article offers insights into the strategies that humans employ when exploring a knowledge network. Readers tend to start from general and popular pages and move progressively to more focused articles at every step. The navigation paths tend to narrow to increasingly more semantically coherent articles (Rodi, Loreto, & Tria, 2017), but by remaining in the semantic neighborhood of the first page. Figure 4b shows the average semantic distance (Piccardi, Gerlach, Arora, & West, A Large-Scale Characterization of How Readers Browse Wikipedia, 2023; Piccardi & West, Crosslingual Topic Modeling with WikiPDA, 2021) from the first article, the entry point to Wikipedia, for each navigation step. Readers diffuse in topic space, moving further from the first article with every step, but significantly slower than a random walker choosing random links on the page. This tendency to keep semantic consistency and not diverge to complete random articles also holds for long sessions up to 100 pageloads when readers fall into a "wiki rabbit hole" (Piccardi, Gerlach, Arora, & West, Going down the Wikipedia Rabbit Hole: Characterizing the Long Tail of Reading Sessions, 2022). Additionally, the average PageRank of loaded articles remains low and even decreases with the session's progress. This observation suggests that readers reach Wikipedia with a latent interest in one topic or a set of topics, and they are not quickly dragged to high centrality nodes. Another property that evolves with the session's progress is the quality of the article loaded. Figure 4c shows the average quality for each navigation step, specifically highlighting two stages. First, the average quality experiences a sharp drop in the second step of the navigation. This behavior can be interpreted as a form of regression to the mean since many sessions start from popular pages with high quality, which thus contribute more to the distribution. By moving one step in the link network, readers naturally reach a page that is, on average, of lower quality. The intuition is confirmed by the behavior of the random walker, which shows the same drop in the first step. Second, the navigation shows a sharp drop in quality with the last pageload, indicating that readers have a higher chance of stopping Wikipedia-internal navigation when reaching a low-quality page. Features correlated with the quality of the article, such as article length and the number of outgoing links, show the same pattern (Piccardi, Gerlach, Arora, & West, A Large-Scale Characterization of How Readers Browse Wikipedia, 2023).

Targeted navigation

Unlike natural navigation, digital traces obtained through wiki games (e.g., Wikispeedia or TheWikiGame) offer insights into how people navigate to find a specific article. The predefined goal represents a clear definition of success, and it enables the understanding of how people find their way to the destination. Participants tend to progress toward the destination in the first part of the exploration by jumping toward high-degree nodes (West & Leskovec, Human Wayfinding in Information Networks, 2012; Helic, 2012). These articles act as hubs of the network and maximize the probability of finding a page closer to the target. Once a hub is reached, people advance to the destination using content features and traverse the semantic space with ever smaller step sizes. These features can predict the destination of the search (West & Leskovec, Human Wayfinding in Information Networks, 2012), with important implications for the design of tools that can assist people in reaching the desired content. These navigation strategies make humans very efficient in finding the shortest paths between two concepts on a knowledge network. Interestingly, this high performance does not necessarily require background knowledge on the topic: simple automatic agents relying on basic features of the articles have performance comparable to humans (West & Leskovec, Automatic Versus Human Navigation in Information Networks, 2012). However, individuals' performance when navigating a knowledge space is influenced by personal characteristics, with younger and multilingual individuals demonstrating higher navigation abilities (Zhu, Yasseri, & Kertész, 2024).

Another advantage of a clear termination state is the possibility to model how people drift away from the best path and understand when users will abandon the exploration. Like in the case of successful navigation sequences, people tend to move quickly to high-degree nodes, but a progressive increase of the semantic distance from the target indicates that the user has lost the right track, and out of frustration, the navigation will be interrupted soon after (Scaria, Philip, West, & Leskovec, 2014). Using properties of the articles and sequential models, the success of a navigation game is predictable (Koopmann, Dallmann, Hettinger, Niebler, & Hotho, 2019).

The paths obtained through targeted navigation gave researchers valuable insights into how humans navigate information networks, but it does not necessarily represent how readers navigate Wikipedia in a natural setup. As we saw in the previous section, natural navigation, defined as sequences of internal clicks, tends to be short, mostly composed of a single pageload.

## 4. How readers leave Wikipedia

Besides an extended network of internal links, Wikipedia contains many links to external resources. External links enrich articles with additional content that should not or cannot be included in Wikipedia itself. There are various reasons to add external links, with linked content ranging from official websites to news articles used as references, to copyrighted material. Articles contain external links that are organized mainly in three different page areas: infoboxes, article bodies, and reference sections. In 2019, English Wikipedia had more than 60M links to external resources, which, over the course of four weeks, received more than 43M clicks (Piccardi, Redi, Colavizza, & West, On the Value of Wikipedia as a Gateway to the Web, 2021). The links in the infoboxes have the highest level of engagement. The per-link click-through rate of links in this area is one click every 110 impressions. For links in article bodies, it is one click every 720 impressions, and for links in reference sections, it is less than one click every 3,000 impressions. In the following, we focus on external links in reference sections and infoboxes, respectively.

Engagement with links in reference sections

References are one of the crucial features of Wikipedia for the content integrity of the platform, and they ensure that the information available in the encyclopedia is externally verifiable. They play an important role in disseminating scientific material (Thompson & Hanley, 2018) and fact-checking (Chen & Roth, 2012), including fact-checking of medical content (Maggio, Steinberg, Piccardi, & Willinsky, 2020; Maggio, et al., 2019). In 2019, Wikipedia contained more than 35M links in the reference section, which, in one month, received more than 14M clicks (Piccardi, Redi, Colavizza, & West, Quantifying engagement with citations on Wikipedia, 2020). Most of the traffic from Wikipedia is directed to archive.org (882K clicks), hosting the Internet Archive's Wayback Machine, a popular service for permanently archiving Web pages. Extracting original URLs from the URLs of archiving platforms reveals that the most clicked domain is google.com, with a significant proportion of clicks going to books.google.com, Google's archive of digitized books. The second most clicked domain is doi.org, the domain for all scholarly articles, reports, and datasets recorded with a Digital Object Identifier (DOI), followed by (mostly liberal) newspapers (The New York Times, The Guardian) and broadcasting channels (BBC).

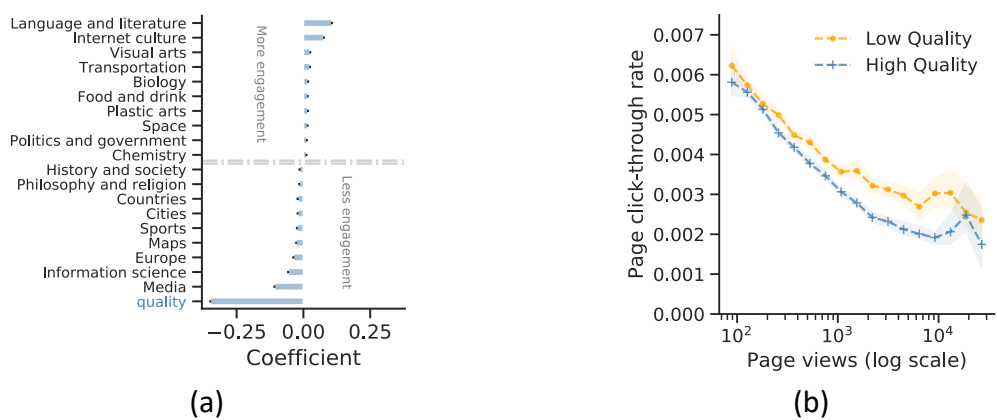

(a)  (b)

*Figure 5.* (a) Coefficients of a logistic regression model predicting high or low engagement on the references of an article ("Language and literature" includes biographies). (b) Comparison of page-specific click-through rate for short vs. long articles, as a function of article popularity. Error bands: bootstrapped 95% CIs. (Figures borrowed from (Piccardi, Redi, Colavizza, & West, Quantifying engagement with citations on Wikipedia, 2020))

Clicks on at least one reference on a visited page happen around once every 340 pageloads. However, there are substantial differences between desktop and mobile: on desktop devices, one in 178 pageloads experience clicks on a reference, over four times as many as on mobile devices, where it happens only once every 769 pageloads.

Readers tend to interact more with the references in articles about biographies, literature, and Internet culture. Figure 5a shows the page features that strongly predict higher and lower engagement with the article's references. Along with topics such as Media and Information Science, the most important negative predictor is the article's quality, which plays a crucial role in the level of engagement. Readers are inclined to seek content beyond Wikipedia by engaging more with citations in lower-quality content that likely does not satisfy their information needs. At the same time, engagement decreases with the page's popularity: high-quality articles tend to be more popular and accessed by a broad and unfocused audience, whereas less popular articles are visited with a specific information need in mind (Singer, et al., 2017). Figure 5b shows that high- and low-quality articles maintain a gap in the page click-through rate (Piccardi, Redi, Colavizza, & West, Quantifying engagement with citations on Wikipedia, 2020) for different levels of popularity.

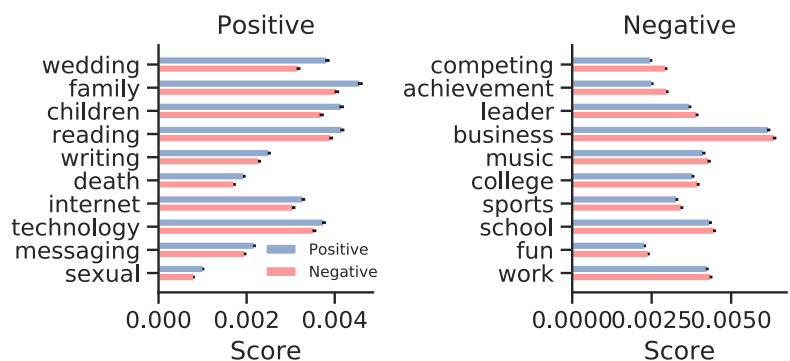

*Figure 6.* Topics most strongly positively or negatively associated with reference clicks. *(Figures borrowed from* (Piccardi, Redi, Colavizza, & West, Quantifying engagement with citations on Wikipedia, 2020)*)*

It has also been found that readers tend to interact more with links of references related to social and life events and relationships. Words such as "dies", "obituary", "married", "wife", "relationship", "sex", "daughter", or "family" appearing in the descriptive text associated with a reference are strong predictors of a click. At the same time, people interact more with references with recent dates, suggesting higher interest in current events. On the other hand, career-related references containing words such as "awards", "debut", or "worked" are less likely to be clicked. Figure 6 shows the topics associated with high interaction with reference links. Human factors such as weddings, family, sex, and death are the most prominent, whereas career-related topics such as competitions or achievements receive less attention.

Engagement with official links in infoboxes

While most external links (95.5%) appear in article bodies and reference sections, a disproportionately large fraction (23%) of the total traffic comes from a relatively small fraction (0.8%) of all external links, namely from official links present in infoboxes (Piccardi, Redi, Colavizza, & West, On the Value of Wikipedia as a Gateway to the Web, 2021). These links typically point to the homepage of the person, organization, or business described in the article. Here, too, readers show different levels of interaction based on the topic of the article.

Official links associated with articles about businesses, educational institutions, and websites have the highest level of engagement. In contrast, official links in articles about geographical content, television, and music exhibit the lowest click-through rates. On average, when the external link is present, the click-through rate is around 2.4%, but with some extreme outliers with a click-through rate of up to 40% and more. For these outliers, Wikipedia serves as a steppingstone between search engines and third-party websites. Typically, these links are URLs that are downranked or censored by search engines, thus not retrievable via search. However, they can often be found in Wikipedia infoboxes, leading search users to detour via Wikipedia. This observation suggests that Wikipedia regularly and systematically meets navigational needs that search engines do not meet, which further confirms Wikipedia's central role in the Web ecosystem.

**The economic value of the outgoing traffic.** Wikipedia is free, and it runs thanks to the donations of thousands of people. Although it is not realistic to evaluate how much money Wikipedia could earn by charging a fee for external clicks, it is possible to approach the question from a different angle, by asking how much money external-website owners would have to pay to obtain an equivalent number of clicks by other means, such as paid ads. An evaluation (Piccardi, Redi, Colavizza, & West, On the Value of Wikipedia as a Gateway to the Web, 2021) based on the cost-per-click estimated via the Google Ads platform showed that the owners of external websites linked from Wikipedia infoboxes would need to collectively pay a total of around $7-13 million per month (or $84-156 million per year) to obtain the same volume of traffic as they receive from Wikipedia for free. The websites that generate the traffic with higher estimated click costs are associated with articles about Business and Biographies. Although this analysis of monetary value should mostly be taken as an indicative back-of-the-envelope calculation, it highlights the importance of Wikipedia as a source of information and as a gratuitous provider of economic wealth.

## 5. Conclusion and research opportunities

Information-seeking is an essential behavioral process that allows people to learn, make decisions, and make sense of the surrounding world. Uncovering the dynamics that guide people in finding information has immediate implications for better understanding our cognitive processes and designing systems that can better accommodate our needs. This chapter offers an overview of the patterns of online knowledge consumption by focusing on the case of Wikipedia, the largest encyclopedia ever created.

Wikipedia fulfills a diverse set of needs that vary across numerous features of the readers, including temporal and geographical properties. Reader behavior shows peculiarities specific to the topics consumed on the platform. Sessions starting from articles about entertainment and biographies tend to be longer and broader compared to sessions starting from articles about STEM topics. This observation suggests a less focused reading intent in articles about entertainment that may be more often connected with exploration driven by boredom (Lemmerich, Sáez-Trumper, West, & Zia, 2019). Similarly, readers manifest the human social nature by engaging more with the citations in biographies and with content associated with human factors such as relationships (wife, family, daughter) and life events (wedding, birth, death). On the contrary, official links receive more engagement on pages about business and education. The platform acts as a gateway for business websites and content that is typically not easily accessible from search engines, and the estimated economic value of this outgoing traffic is on the scale of several million dollars per month. This observation, combined with the frequent click of references routing the readers to open access documents, highlights the previously largely undocumented and underestimated navigation role of Wikipedia for scientific and business content.

The quality of the articles impacts navigation behavior, and the exploration of a path by following internal links has a higher chance of terminating in low-quality articles. This observation is aligned with the definition of information scent used in information foraging theory. Humans hunting for information follow the scent with higher chances of leading to the desired content; when the scent loses intensity, they move to other, more promising information sources. Further evidence of this hypothesis is that low-quality articles exhibit higher engagement with the citations suggesting that unsatisfied readers abandon the platform to satisfy their information needs somewhere else.

**Research opportunities.** Wikipedia is an intricate system composed of linked concepts in multiple languages, multimedia items, and human behavioral traces. The richness of its data enables scholars to address numerous important research questions. Investigating the patterns associated with content consumption has implications for developing theoretical frameworks to describe our navigation patterns. Understanding what drives the readers to follow specific trails can inform researchers about the specific properties of the information scent that guide our search for information online (Chi, Pirolli, & Pitkow, 2000). These findings can be instrumental in developing novel theories on how humans move in information networks. Future research in this direction could combine the current understanding of knowledge consumption patterns, as sketched in this chapter, with qualitative research to connect the behaviors observed with the readers' intent. Progress in this direction can guide future developments of the Wikipedia platform. By adding the readers' behavior in the design loop, developers and user experience designers can implement new forms of knowledge access. Modeling how readers consume the content to infer their intention can—potentially with the support of AI—help augment the navigation. In the future, adaptive interfaces optimized for different knowledge consumption patterns can be employed to offer customized Web experiences. Similarly, recommender systems for the editor community can exploit signals obtained from reader behavior as a hint on how to improve the content. Frequent sequences of articles can be used to infer learning pathways or improve the connectivity of the links network, and content with high engagement (i.e., references without links) can be recommended for improvement.

Given the large scale of Wikipedia, the platform also represents a unique dataset to measure global sociocultural differences and the attention response to external events. For example, socioeconomic conditions may be associated with diverse information needs that influence reading patterns. Most current work on reader behavior is focused on the English edition of Wikipedia, and expanding to multiple languages can give us insights into common behavior shared across languages. Finally, combining cross-topic and cross-lingual studies (Piccardi & West, Crosslingual Topic Modeling with WikiPDA, 2021; Josifoski, Paskov, Paskov, Jaggi, & West, 2019) with readership patterns can expose potential biases to which readers are unknowingly exposed.

In conclusion, Wikipedia is an invaluable resource for studying and modeling knowledge consumption patterns and human behavior on the Web. Since our understanding is still far from complete and a large portion of the data—in particular the clickstream (Arora, Gerlach, Piccardi, García-Durán, & West, 2022)—is public, the opportunities for the research community motivated to understand human behavior are enormous.